\documentclass[%
 reprint,
superscriptaddress,
 amsmath,amssymb,
 aps,
]{revtex4-2}

\usepackage{graphicx}
\usepackage{dcolumn}
\usepackage{bm}
\usepackage[colorlinks,allcolors=blue]{hyperref}

\begin{document}

\preprint{APS/123-QED}

\title{X-ray pulse generation with ultra-fast flipping of its orbital angular momentum} 

\author{J. Morgan}
 \email{jmorgan@slac.stanford.edu}
 \affiliation{SLAC National Accelerator Laboratory, Menlo Park, California, 94025, USA}

\author{B. W. J. M$^{\rm c}$Neil}
 \affiliation{University of Strathclyde (SUPA), Glasgow G4 0NG, United Kingdom}
 \affiliation{Cockcroft Institute, Warrington, WA4 4AD, UK}
\affiliation{ASTeC, STFC Daresbury Laboratory, Warrington, WA4 4AD, UK}

\date{\today}

\begin{abstract}

A method to temporally tailor the properties of X-ray radiation carrying Orbital Angular Momentum (OAM) is presented. In simulations, an electron beam is prepared with a temporally modulated micro-bunching structure which, when radiating at the second harmonic in a helical undulator, generates OAM light with a  corresponding temporally modulated intensity. This method is shown to generate attosecond pulse trains of OAM light without the need for any additional external optics, making the wavelength range tunable. In addition to the OAM pulse train, the method can be adapted to generate radiation where the handedness of the OAM mode may also be temporally modulated (flipped).

\end{abstract}

\maketitle

\section{\label{sec:level1}Introduction}

Light-matter interactions play an important role in our understanding of material interactions across most of science, from how viruses function to the properties of matter at the centre of gas giant planets. The material properties that can be observed depend critically upon the form of the light used: wavelength, intensity, temporal duration etc. Much research is therefore devoted to improving upon, or developing new forms of light, that can access previously unobserved material properties.

Increasing attention is being given to light with helical wave-fronts, as its transverse phase rotation means the light carries orbital angular momentum (OAM). This OAM is distinct from the  spin momentum of the light and can be significantly greater, allowing another dimension of light-matter interactions to be explored and utilised. Allen \textit{et al.}~\cite{allen1992orbital} was first to recognise OAM is a property belonging to light with an azimuthal phase dependence given by $\exp{i\ell\phi}$, where $\phi$ is the azimuthal angle and $\ell$ the mode index. 
Since publication of this seminal paper in 1992, there has been significant research interest in the scientific applications of this light which has benefited numerous fields~\cite{padgett2017orbital} such as imaging~\cite{https://doi.org/10.1002/lpor.200900047, Furhapter:05} and communication~\cite{Willner:15}. As OAM can be exchanged with atoms and molecules, there has also been great interest in the study of OAM light with matter~\cite{sym13081368, PhysRevLett.96.243001, PhysRevLett.89.143601, PhysRevA.83.065801,NJP:2010}. This is being further developed by investigating new methods to generate OAM light pulses that can interact with matter at its fundamental spatial (X-ray wavelengths) and temporal (attosecond) scales \cite{Lloyd_Hughes_2021, Shi_2020, Biegert_2021, PhysRevLett.126.117201}.


There has been specific interest in the generation of pulses of OAM light at XUV and X-ray wavelengths due to predicted experimental applications~\cite{PhysRevLett.98.157401,  patchkovskii2012high}. To meet this need, short individual pulses and trains of pulses have been generated at XUV using high harmonic generation~\cite{dorney2019controlling, PhysRevLett.113.153901,PhysRevLett.111.083602,wang2019intense}. Although these methods are predicted to work into the soft X-ray, this is yet to be demonstrated. 
OAM light can be generated by manipulating the phase profile of Gaussian beams with $l=0$ using optical elements which convert the light phase. These   require precise fabrication which can be particularly challenging at shorter wavelengths, although elements which work at XUV and X-ray wavelengths are becoming increasingly available~\cite{PhysRevSTAB.13.030701}. However, these optical elements are not tuneable, so different elements are required at different wavelengths.  Such conventional optical elements can also pose a barrier to the rate at which the OAM mode of the light can be controlled. Routine OAM switching optics, such as spatial light modulators, only achieve a mode switching rate in the kHz range, although this could possibly be improved on with more complex techniques. 

In recent developments at short wavelengths, a temporal variation in the OAM of a light beam (a `self-torqued' beam) has been achieved using harmonic generation~\cite{rego2019generation} and opens up another new branch of potential applications.

Free Electron Lasers (FELs) provide a method for generating intense OAM light without the need for additional optical elements. There has been significant progress in the generation of light with OAM in a FEL including experimental demonstration~\cite{PhysRevX.7.031036,Hemsing2013} and the proposed combination of OAM modes to generate Poincar\'e X-ray beams~\cite{Morgan_2020}. However, current FEL studies have not yet considered the detailed control of the temporal properties of the OAM light. 

In this article, a new method is proposed to use a FEL to generate high power, X-ray OAM light with attosecond intensity modulation. The approach uses a FEL `afterburner' configuration~\cite{PhysRevSTAB.13.030701}. This consists of a main FEL lasing section that first develops a strongly modulated electron beam micro-bunching which is then injected into  the helical undulator afterburner where the attosecond timescale, temporally modulated OAM light is generated.  A small modification to this approach, the addition of a second afterburner undulator, allows the generation of light pulses which alternate between $l = \pm 1$ OAM modes, again at attosecond timescales. The alternation of OAM light at such fast timescales and short wavelengths would be the first time such output would be available for users.



\section{Orbital Angular Momentum From Relativistic Electron Beams}


OAM was shown to be carried by the harmonic radiation emitted from electrons propagating in a spiral path through a helical undulator~\cite{sasaki2008proposal, SASAKI200743}. The light is characterised by the OAM index:
\begin{equation}
 \ell=\pm(h-1).
\end{equation}
Here, $h$ is the harmonic number and the mode index $\ell$, is sign dependent upon the handedness of the undulator. This theoretical work was subsequently experimentally corroborated with OAM demonstrated to be carried by both incoherent and coherent undulator radiation~\cite{PhysRevX.7.031036, PhysRevLett.113.134803}. 
Many methods have been proposed to generate OAM light with relativistic electron beams. Some methods do not require emission at the harmonics~\cite{Hemsing2013,PhysRevLett.112.203602} while another method is to simply filter out the harmonic OAM emission from a helical undulator.

Here, a FEL afterburner arrangement is used ~\cite{Hemsing2013,PhysRevLett.110.104801} in which a FEL amplifier first bunches the electron beam before it is propagated into a downstream helical undulator, the afterburner, where the OAM light emission takes place. The pre-prepared electron bunching is at the second harmonic of the helical undulator afterburner resonance and generates the coherent harmonic emission of radiation which carries the desired OAM.  
 

\section{Generation of light pulses with alternating OAM}


A method is now proposed to generate a train of ultra-short light pulses with alternating OAM of $\ell=+1, -1, +1,...$. The method uses a similar setup as the 2nd harmonic helical undulator OAM afterburner, but with an electron beam prepared so that its micro-bunching is modulated into a periodic comb structure. This can be obtained by using an energy modulated electron beam in the first FEL amplifier that induces the electron micro-bunching. When emitting in a short helical afterburner undulator, the modulated micro-bunching of the electron beam causes emission of a train of coherent OAM light pulses spaced at the modulation frequency, in a  similar way to that described in~\cite{PhysRevLett.110.104801}. If a second OAM afterburner undulator, with the same tuning but opposite handedness as the first, is added, a second interweaving light pulse train can be superimposed so that the final light generated alternates between OAM mode pulses of approximately the same power.

\subsection{Modulating the electron beam micro-bunching \label{SectMod}}

When a relativistic electron beam is prepared with a sinusoidal energy modulation, $\gamma(t)=\gamma_0+\gamma_m\cos({\omega_m t})$, the energy gradients about the mean energy $\gamma_0$ `spoil' the FEL interaction there. No seeding of the FEL is required and the process starts up from noise.  If the modulation amplitude, $\gamma_m$, is sufficiently large,  the only regions that develop any significant electron FEL micro-bunching are about the minima of the energy modulated beam~\cite{PhysRevLett.110.104801}. After a preparatory process in a FEL amplifier, the energy modulated electron beam  has regions of relatively high micro-bunching separated by the modulation period $\lambda_m=2\pi / \omega_m$. 

When the electron beam with modulated electron micro-bunching structure is propagated through a short OAM afterburner undulator module, the power of the coherent OAM light it radiates has the same modulation period. 


The modulation in the temporal power profile of the radiation corresponds to a frequency spectrum which is broader than the typical FEL output and is discretized into frequency modes. To examine the effect of this on OAM mode purity we first present analysis of the spectral output of a single undulator module. Following this, simulations of the method to generate pulse trains of OAM light, as well as pulse trains with alternating OAM, are presented.

\subsection{\label{section: single undulator spectrum} Spontaneous emission spectrum of a single undulator module}

\begin{figure}[htb]
    \centering
    \includegraphics[width=0.2\textwidth]{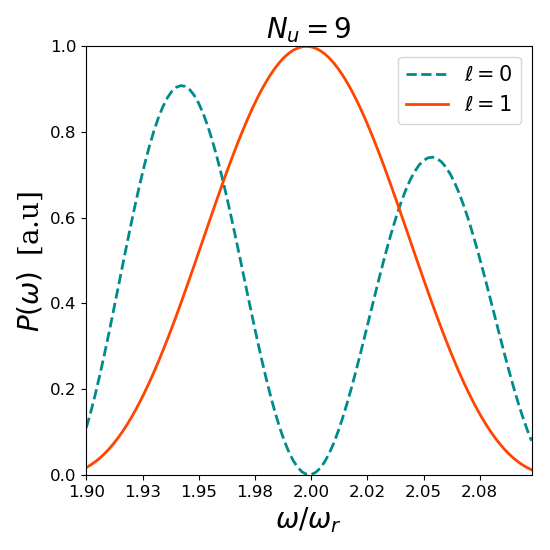}
    \includegraphics[width=0.2\textwidth]{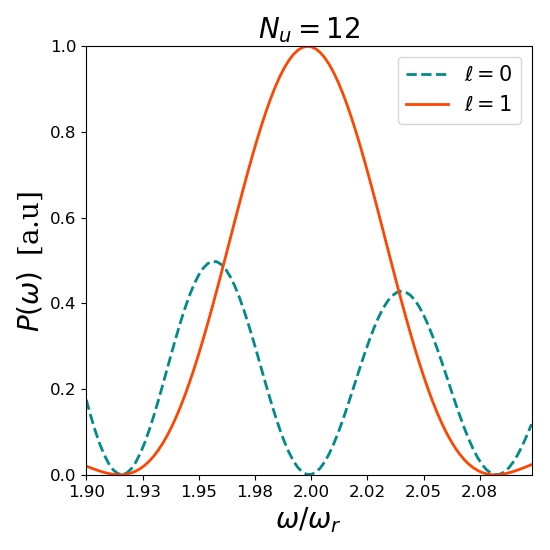}
    \includegraphics[width=0.2\textwidth]{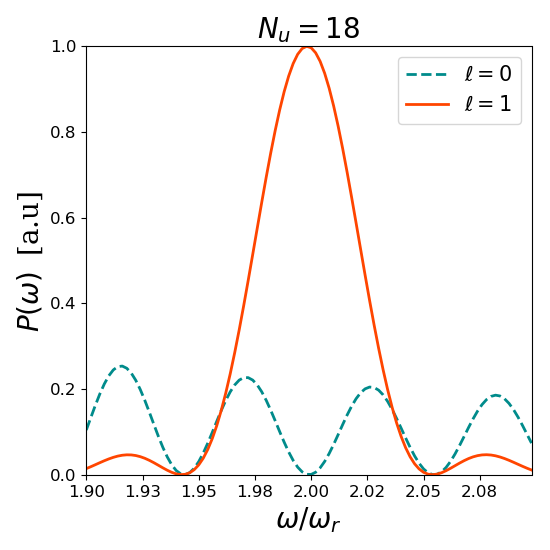}
    \includegraphics[width=0.2\textwidth]{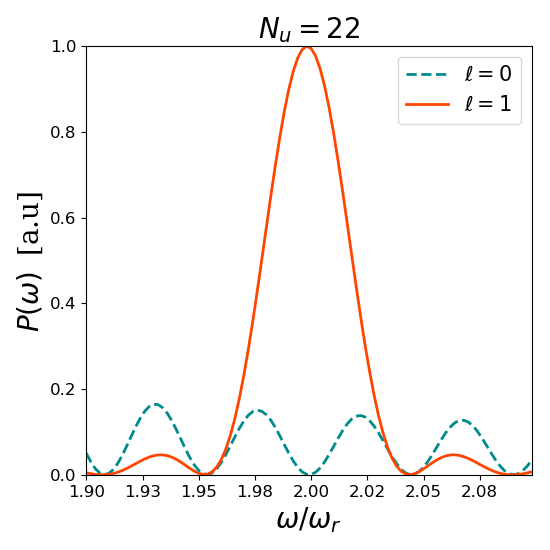}
    \caption{Spectrum of helical undulator radiation for different number of undulator periods $N_u = 9, 12, 18, 22$ with $a_u=2.634$, $\lambda_u=3.9$cm, $\gamma=7869$ and $\sigma_x=26 \mathrm{\mu m}$.}
    \label{Spectrum for different Nu}
\end{figure}

To examine the OAM content of a helical undulator at the second harmonic, the far field radiation emitted by a single electron propagating in a helical undulator~\cite{PhysRevAccelBeams.23.020703} is multiplied by the term $\exp (-\sigma_x^2 \omega^2 / c^2\sin^2\theta)$, which captures the restriction imposed by the beam's transverse dimensions on the angle of emission $\theta$. A Gaussian transverse profile is assumed, where $\sigma_x$ is the rms beam radius. The radiation spectrum is then considered in terms of OAM content. Fig.~\ref{Spectrum for different Nu} plots the radiation spectrum decomposed into different OAM modes (as in~\cite{Hemsing2013}) for different number of undulator periods, $N_u$. An undulator parameter of $a_u>1$ is used, increasing the fractional power emitted into the $\ell=1$ mode~\cite{PhysRevAccelBeams.23.020703}.

It can be seen that undulators with a lower number of periods $N_u$,  have significant emission into the fundamental Gaussian mode with $\ell=0$. For a long electron beam bunched uniformly at the second harmonic, the bunching picks out a narrow region of the FEL bandwidth about the peak of the $\ell=1$ emission. There is therefore little emission into the Gaussian mode when the micro-bunching structure is not modulated as assumed in the simple OAM afterburner.


When the electron beam micro-bunching is modulated to have a comb structure, as described in Section \ref{SectMod}, then radiation pulses are generated at the modulation period. For a small number of undulator periods $N_u$, it can be seen from Fig.~\ref{Spectrum for different Nu} that a significant fraction of the power can be emitted off-resonance into the $\ell=0$ mode. Increasing $N_u$ both increases the relative fraction of power in the $\ell=1$ mode, and decreases the  bandwidth with FWHM spectral width of harmonic, $h$ given by:
\begin{equation}\label{FWHMfrequency}
   \Delta \omega_{h}=\frac{0.88}{hN_u}\omega_h(\theta),
\end{equation}
where $\omega_h(\theta)$ is the peak of the spectrum.




\section{Numerical Simulations}

\begin{figure}[htb]
    \vspace{0.05cm}
    \centering
    \includegraphics[width=0.5\textwidth]{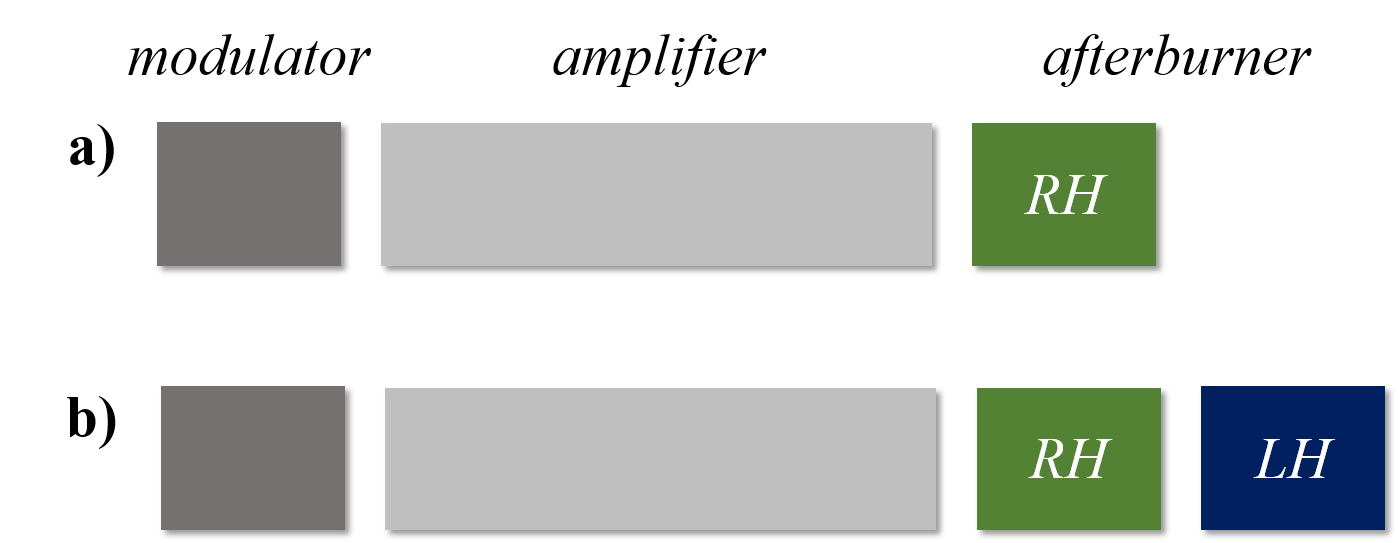}
    \caption{
    The electron's energy is modulated before injection into an amplifier undulator where the electron beam develops a similarly modulated micro-bunching structure. A pulse train of radiation carrying OAM radiation is then emitted in the short helical afterburner tuned to the second harmonic of the electron micro-bunching. In $a)$ a single helical undulator afterburner will produce a single handed OAM pulse train, while in $b)$ a second undulator with opposite handedness is added. The pulse trains emitted within each afterburner can then be made to alternate spatially, resulting in the generation of OAM pulses that alternate between $\ell=\pm 1$. }
    \label{fig:schematic}
\end{figure}

Using the FEL simulation code Puffin~\cite{campbell2012puffin, puffinUpdate}, the generation of OAM pulse trains in the soft X-ray is demonstrated. Fig \ref{fig:schematic} shows a schematic of the layout used to generate two different types of pulse trains: a) each pulse has an OAM with index of $\ell=1$; and b)  the pulses have an OAM index that alternates between $\ell=\pm 1$. 

The method is modeled using parameters inspired by the LCLS-II project at SLAC~\cite{LCLSIIScience}. A $40$ fs long electron beam is considered with a flat top beam current profile and current $I=1$~kA. The energy is $4\ \mathrm{GeV}$ and the radius and emittance are  $\sigma_x=22\ \mathrm{\mu m}$ and $\epsilon _{x,y}= 3.5\  \mathrm{mm-mrad}$ respectively. 



\subsection{OAM pulse train generation}
The electron beam is first prepared in a seeded undulator modulator to give it an energy modulation of amplitude ${\gamma_m/\gamma_0}=0.0014\approx \rho$ (where $\rho$ is the  FEL parameter ~\cite{SASE})
 and period  $\lambda_m= 100\ \mathrm{nm}$. This relatively long wavelength energy modulation is imposed as initial conditions on the electron beam without the need for simulations (it is relatively easy to achieve such modulation as detailed in~\cite{starttoend, doi:10.1063/1.3605027}.) 

\begin{figure}[htb]
    \vspace{0.75cm}
    \centering
    \includegraphics[width=0.5\textwidth]{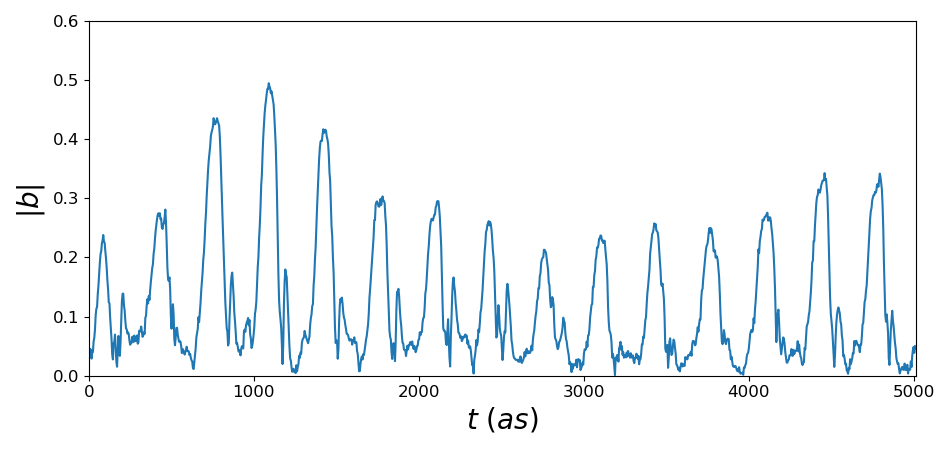}
    \caption{Electron bunching parameter ($0 < |b| < 1$) as a function of relative time for a section of electron beam captured near the end of the FEL amplifier.}
    \label{fig:Bunching comb}
    
\end{figure}

The energy modulated electron beam is then micro-bunched in a planar undulator FEL amplifier with rms undulator parameter $\bar{a}_u=1.72$, period $\lambda_u =3.9$~cm and at the resonant wavelength $\lambda_r=1.25\ \mathrm{nm}$. The micro-bunching at the resonant wavelength occurs at the minima of the energy modulations where the gain is greater. Fig.~\ref{fig:Bunching comb} shows the bunching structure which has been established at the end of the FEL amplifier. The regions of the electron beam with high electron bunching will emit coherently in the afterburner so that a modulation in the electron bunching parameter leads to a corresponding modulation in the  radiation intensity.

The first amplifier stage has a small reverse taper in the undulator parameter $\bar{a}_u$~\cite{PhysRevSTAB.16.110702}. This reduces the intensity of radiation emitted by the electrons without any significant reduction to the electron bunching at $\lambda_r$. In addition to using a reverse taper to suppress the radiation output from the modulator section, a small steering of the electron beam away from it before entry into the afterburner sections can also be used.  
In the simulations presented here, the steering of the electron beam is not modeled explicitly - the radiation generated in the modulator is simply removed before injection into the afterburner sections.

Following the preparation of the $4$~GeV electron beam, with a  micro-bunching at $1.25$~nm modulated at $\lambda_m=100$ nm, it is then injected into the afterburner section where the OAM output is generated. The first undulator module is a right-handed helical undulator with $\bar{a}_u=2.634$, $\lambda_u=3.9$~cm and $N_u=20$. The fundamental resonant wavelength is then $\lambda_1=2.5$~nm, so that the electrons are micro-bunched at the 2nd harmonic. The relative propagation distance, or `slippage', of a wavefront through the electrons is then $N_u\lambda_1=\lambda_m/2$, one half of the electron beam modulation period.

\begin{figure}[htb]
    \vspace{0.75cm}
    \centering
    \includegraphics[width=0.5\textwidth]{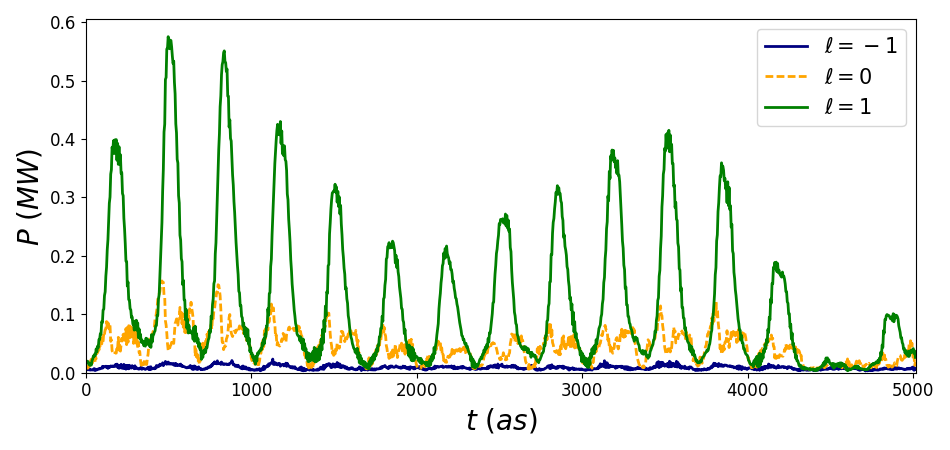}
    \caption{Temporal plot of the radiation power output from the $\ell=1$ helical afterburner decomposed into OAM modes. The radiation wavelength $\lambda_{OAM}\approx 1.25$~nm and the pulse FWHM durations are $\tau \approx 40\; \lambda_{OAM}/c=167$~as.}
    \label{fig:OAM_pulse_train}
    
\end{figure}

In Fig.~\ref{fig:OAM_pulse_train} it is seen that the modulated micro-bunching of the electron beam at the 2nd harmonic of resonant wavelength on the helical afterburner emits a pulse train of OAM light of mode $\ell=1$ and wavelength $\lambda_{OAM}=1.25$~nm. Each pulse has a FWHM pulse duration of $\tau\approx 167$~as, approximately 40 optical cycles corresponding to twice the number of helical undulator periods. 

It is seen that in addition to the OAM emission there is also some coherent emission into the $\ell=0$ mode. This emission is expected due to the relatively short length of the afterburner undulators needed to produce short pulses as discussed in section \ref{section: single undulator spectrum}. The length of the undulator is fixed to generate  a specific pulse length. However, reducing the crossectional radius of the electron beam is one method that will increase the emission into the OAM mode.

\subsection{Time Varying OAM}

A left-handed helical undulator is now added to the right-handed afterburner (see Fig~\ref{fig:schematic}~b)) with the same length and tuning. Previous versions of this crossed undulator configuration have been used to enable polarization control at the fundamental~\cite{ferrari2019free, deng2014polarization}, spatially varying polarization~\cite{PhysRevAccelBeams.22.110705} and the generation of Poincar\'e beams~\cite{Morgan_2020}. Recently, crossed undulators have been proposed as a method to modulate the polarization much in the same way we alternate the OAM here~\cite{PhysRevAccelBeams.24.010701, PhysRevAccelBeams.23.120701}. 

\begin{figure}[htb]
    \centering
    \includegraphics[width=0.5\textwidth]{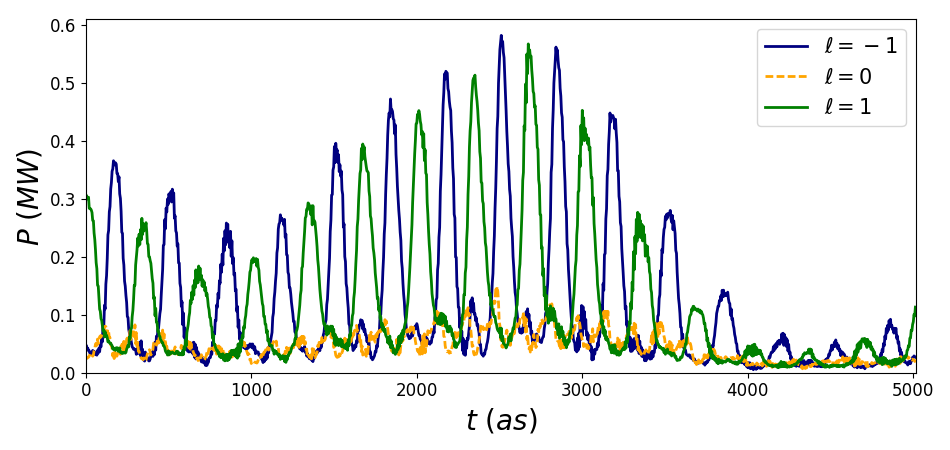}
    \includegraphics[width=0.5\textwidth]{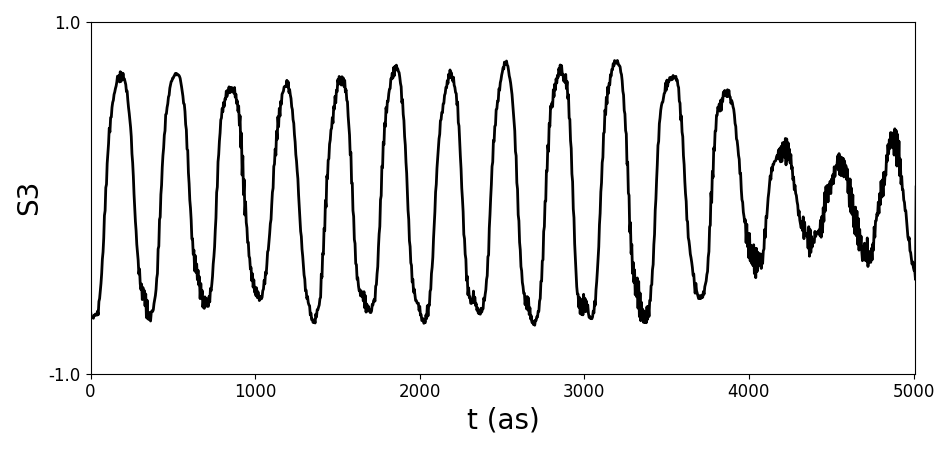}
    \caption{(top) Power Vs relative time at the end of the afterburner for FEL radiation decomposed into OAM different modes.(bottom) Stokes parameter $s_3$ vs time.}
    \label{fig:alternating_OAM}
\end{figure}

The additional undulator emits a second pulse train with OAM mode $\ell=-1$. The slippage in the first undulator means that this $\ell=-1$ pulse train is shifted relative to the first by half a modulation period.  The combined pulse train from the two afterburners then consists of a sequence  of pulses that alternate between $\ell=1$ left-hand circularly polarized light  and $\ell=-1$ and right-hand circularly polarized light. Fig.~\ref{fig:alternating_OAM} shows the power radiation pulse train after the second afterburner decomposed into the different OAM modes. The individual pulses have the same pulse length, $\tau_p=167$, as before. The time between pulse peaks is also $167$ as, corresponding to a switching rate of $6$~PHz. 

\begin{figure}[htb]
    \centering
    \begin{minipage}{0.24\textwidth}
    \includegraphics[width=\textwidth]{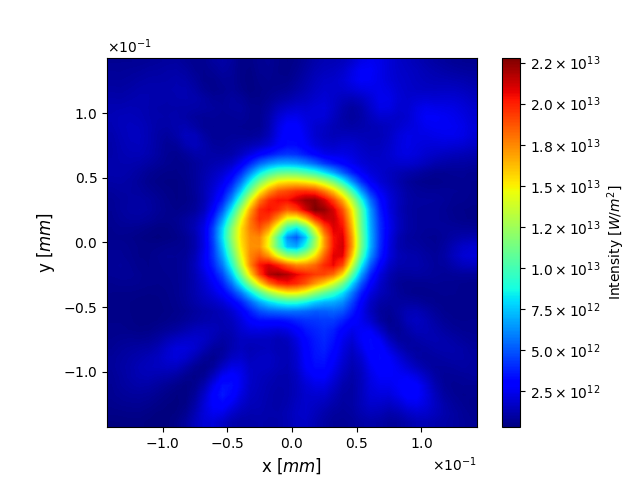}
    \end{minipage}\hfill
    \begin{minipage}{0.24\textwidth}
        \includegraphics[width=\textwidth]{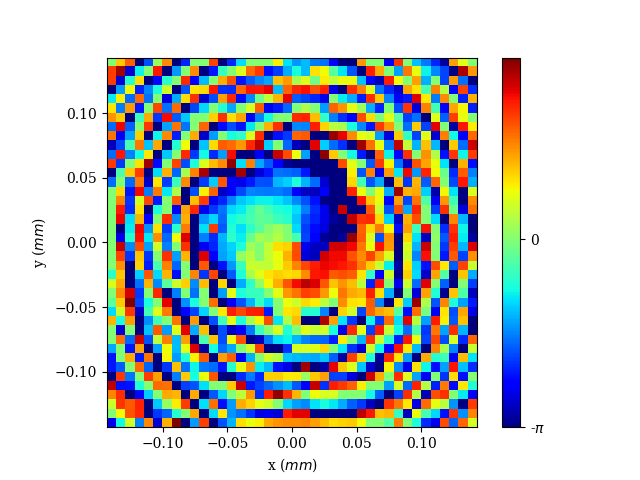}
       \end{minipage}\hfill
    \begin{minipage}{0.24\textwidth}
    \includegraphics[width=\textwidth]{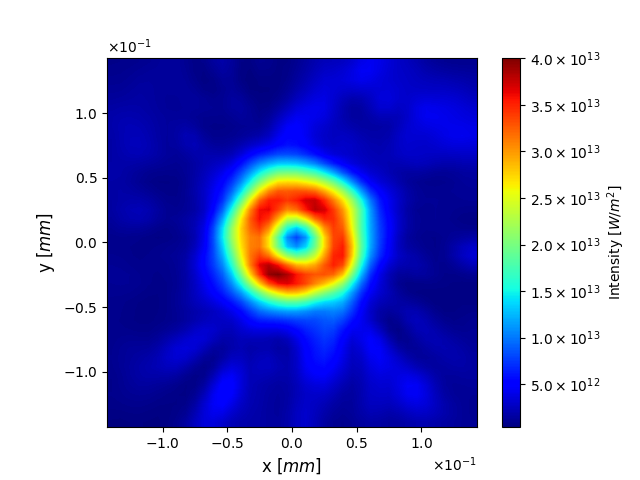}
    \end{minipage}\hfill
    \begin{minipage}{0.24\textwidth}
        \includegraphics[width=\textwidth]{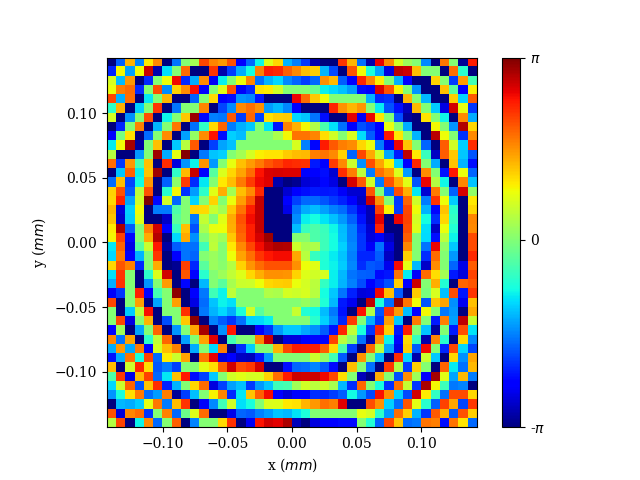}
           \end{minipage}\hfill
    \caption{(Left) Pulse intensity averaged over a single pulse and (right) instantaneous phase for the pulses found as relative times (top) $t=2516$ as and (bottom) $t=2683$ as in Fig~\ref{fig:alternating_OAM}.  }

    \label{fig:alternating_OAM_phase}
\end{figure}

The intensity profile and phase of the radiation of an $\ell=1$ and $\ell=-1$ is shown in Fig~\ref{fig:alternating_OAM_phase}. The intensity (averaged over a single spike) shows the typical OAM `doughnut' profile in the transverse plane. The phase of the radiation is captured near the intensity peaks of two neighboring radiation spikes. There is a clear azimuthal phase distribution which changes handedness in the $167$~as interval between snapshots. 

Also included in the Fig.~\ref{fig:alternating_OAM} is the normalized Stokes parameter for circular polarization

\begin{equation}
    s_3 = \frac{|E_R|^2-|E_L|^2}{|E_R|^2+|E_L|^2} ,
\end{equation}

The polarization of the light alternates along with the OAM mode. This then provides an interesting case as the polarization rotates in the same direction as the OAM. The total angular momentum contribution from the OAM plus the spin momentum add and therefore the pulse train alternates between total angular momentum $\pm 2\hbar$. We also note that tuning one of the afterburner undulators so that emission is at the fundamental frequency will produce pulses with no OAM. This $\ell=0$ mode will be emitted from helical and planar undulators and thus a variety of OAM and spin combinations are available. 

If OAM pulses with only one polarization are required, a linear polarizer which will filter out one linear polarization component may be a solution. Filtering out linear polarization corresponding to the polarization of the amplifying FEL section would also help ensure light emitted in this section does not interfere with the final pulse train delivered to experiment. 

\section{Conclusion}

Ultra-fast manipulation of the temporal properties of FEL generated OAM light have been proposed and simulated for the first time. The method can be used to generate a pulse train of light carrying OAM and also alternate the handedness of the OAM at ultra-fast timescales. It was shown that the temporal profile of OAM radiation emitted from electrons can be controlled by tailoring the structure of micro-bunching  in the electron beam. In addition, since the handedness of the OAM is determined by the magnetic undulator and not external optics, FELs are uniquely positioned to offer ultra-fast flipping of OAM handedness at X-ray wavelengths. As the method uses a relatively simple afterburner configuration, it is relatively cost effective to implement at existing FEL facilities.

\section{Acknowledgments}

The authors would like to thank E. Hemsing for providing useful comments in the preparation of this paper.
This work was supported by U.S. Department of Energy Contract No. DE-AC02-76SF00515 and Award No. 2021-SLAC-100732.
We are grateful to funding from the Science and Technology Facilities Council (Agreement Number 4163192 Release\#3); ARCHIEWeSt HPC, EPSRC grant EP/K000586/1; EPSRC Grant EP/M011607/1; John von Neumann Institute for Computing (NIC) on JUROPA at J\"ulich Supercomputing Centre (JSC), project HHH20.

\providecommand{\noopsort}[1]{}\providecommand{\singleletter}[1]{#1}%
\providecommand{\newblock}{}

\end{document}